# Theoretical Study of the Thermal Decomposition of a Jet Fuel Surrogate

*B. Sirjean[1], O. Herbinet[1], P.A. Glaude[1], M.F. Ruiz-Lopez[2] and R. Fournet[1]*

*[1]Départment de Chimie Physique des Réactions, CNRS - Nancy Université, France*

*[2]Equipe de Chimie et Biochimie Théorique, SRSMC, CNRS - Nancy Université, France*

In a scramjet, the fuel can be used to cool down the engine walls. The thermal decomposition of the jet fuel changes the reacting mixture before its combustion. A numerical study of the pyrolysis of norbornane, a jet fuel surrogate, has been performed. Rate constants of some sensitive reaction channels have been calculated by means of quantum chemical calculations at the CBS-QB3 level of theory. The mechanism has been validated against experimental results obtained in a jet-stirred reactor and important and/or sensitive pathways have been derived.

## Introduction

Jet fuels are complex mixtures of hundreds to thousands of hydrocarbons containing a large number of carbon atoms, typically ranging from $C_7$ to $C_{16}$, and belonging to different chemical classes (*n*-alkanes, isoalkanes, cycloalkanes, alkenes, and aromatics) [1]. Including all of these components in a detailed kinetic chemical model is not computationally feasible. For that reason, models with a limited number of components (called surrogate fuel) are used to deal with these practical fuels. If many chemical kinetic studies on the oxidation and the pyrolysis of linear and branched alkanes have been performed in recent years, there are considerably fewer studies on the kinetics of cycloalkanes [2]. The emergence of oil-sand-derived fuels containing a large proportion of cycloalkanes [3] has led recently to an increase of kinetic studies on this class of hydrocarbon (see for example [4] and references therein). However, these studies are focusing only on monocyclic alkanes and more precisely, in almost all cases, on the simplest mono-cycloalkane: cyclohexane. But most of cycloalkanes in fuels are rather large and complex cyclic alkanes and the comprehension of the thermal decomposition of polycyclic alkanes is still far from complete and represents a major kinetic issue.

Polycyclic alkanes can be the major components of many jet fuels. For example, the tricyclic alkane exo-tricyclo[5.2.1.0$^{2,6}$]decane (Figure 1) is the main component of synthetic fuels (e.g., RJ-6, JP-9, and JP-10) that are used in aircraft due to their high volumetric energy content. As these fuels are used in scramjet systems to cool down the engine walls, their thermal decomposition occurs before their introduction into the combustion chamber. Thus a detailed chemical kinetic model of the pyrolysis of the fuel is necessary to estimate the composition of the reacting mixture before its combustion.

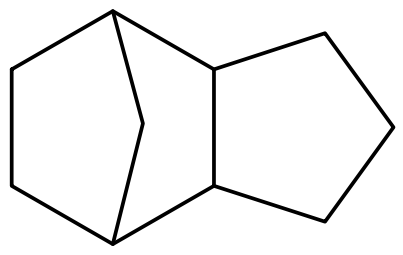

**Figure 1: Structure of exo-tricyclo[5.2.1.0$^{2,6}$]decane ($C_{10}H_{16}$) also called tricyclodecane in the text.**

The thermal decomposition of tricyclodecane has been recently studied by Herbinet et al. (2006) in a jet-stirred reactor [5]. These authors have also proposed a detailed chemical kinetic model to reproduce their experimental results. This model was validated against their data and against the few experimental data available in the literature. Their model has been developed by applying the same systematic method as the EXGAS software [6]. They used bond additivity methods to estimate the thermodynamic data and correlations between structure and reactivity to estimate the kinetic parameters. In their work, Herbinet et al. have showed with sensitive and flow rate analysis that the unimolecular initiations of tricyclodecane play a very important role on the decomposition of tricyclodecane and underlined the lack of thermokinetic data on the reactions of polycyclic alkanes.

Nowadays, accurate thermodynamic and kinetic parameters can be obtained by high-level quantum chemistry methods such as the CBS-QB3 model chemistry proposed by Petersson and co-workers [7]. Unfortunately, applying this level of calculation on tricyclodecane (10 heavy atoms) to describe its unimolecular decomposition reactions is within the limits of the number of heavy atoms that this method can take into account. For that reason, we have chosen to study a model molecule of polycyclic alkanes: bicyclo[2.2.1]heptane also called norbornane (Figure 2).

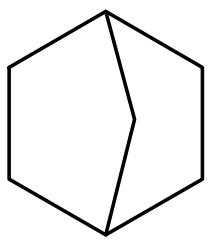

**Figure 2: Structure of bicyclo[2.2.1]heptane ($C_7H_{12}$) also called norbornane**

In a previous paper, [8] we have studied the thermal decomposition of norbornane (dissolved in benzene) in a jet-stirred reactor. A total of 25 reaction products were identified and quantified by gas chromatography, among which the main ones are hydrogen, ethylene, and 1,3-cyclopentadiene. A mechanism investigation of the thermal decomposition of the norbornane-benzene binary mixture has been performed.

In this work, we present a complete model for the pyrolysis of norbornane. The kinetic and thermodynamic data of important reactions, e.g. the unimolecular initiations by C-C bond fission of norbornane, the fate of generated diradicals, the reaction of transfer and propagation of norbornyl radicals, and sensitive reaction pathways, are calculated with the high accuracy CBS-QB3 method. The completeness of the detailed chemical kinetic model is achieved by applying the same systematic method as the EXGAS software [6]. Simulations made with this model are then compared to experimental results.

In this paper, we detail the detailed chemical kinetic model developed and examine the sensitive reaction channel described by quantum chemistry calculation. Finally we compare experimental and computational results.

## 1. Description of the detailed kinetic model

The detailed kinetic model has been constructed by using the same systematic method as EXGAS software, which performs an automatic generation of mechanisms, but cannot be directly applied to polycyclic compounds.

### *a. General features of EXGAS system*

The systematic method leads to a reaction mechanism made of three parts:

- A comprehensive primary mechanism, in which the only molecular reactant considered is the initial organic compound. The primary mechanism to model the pyrolysis of norbornane includes the following elementary steps:
  – Unimolecular initiation steps,
  – Decomposition by β-scission of alkyl radicals,
  – Isomerization of alkyl radicals,
  – H-abstraction reactions from the initial reactants by small radicals.

- A $C_0$–$C_2$ reaction base, including all the reactions involving radicals or molecules containing less than three carbon atoms [17], which is coupled with a reaction base for $C_3$–$C_4$ multi-unsaturated hydrocarbons [18], such as propyne, allene or butadiene, including reactions leading to the formation of benzene and in which pressure-dependent rate constants are considered.
- A lumped secondary mechanism, containing reactions consuming the molecular products of the primary mechanism, which are not included in the reaction bases.

Thermochemical data for molecules or radicals are automatically computed using software THERGAS [19], based on group additivity, and stored as 14 polynomial coefficients, according to the CHEMKIN II formalism [20]. The kinetic data of isomerizations, recombinations and the unimolecular decompositions are calculated using software KINGAS [21], based on thermochemical kinetics methods. The kinetic data, for which the calculation is not possible by KINGAS, are estimated from correlations, which are based on quantitative structure–reactivity relationships and obtained from a literature review. The main features of these calculations and estimations have been summarized in previous descriptions of EXGAS [22].

### *b. Ab initio study of reactions of the primary mechanism*

#### *i. Computational method*

Calculations were performed with Gaussian 03 Rev. B.05 [9]. The composite method CBS-QB3 [7] has been applied for all the species involved in the mechanism. Diradicals species (all in singlet state) have been described with a modified version of the CBS-QB3 method proposed by Sirjean et al. [10]. Evaluation of the vibrational frequencies at the B3LYP/cbsb7 [11-12] level of calculation confirmed that all transition states (TS) have one imaginary frequency. Intrinsic Reaction Coordinate (IRC) calculations have been systematically performed at the B3LYP/6-31G(d) level on transition states, to ensure that they are correctly connected to the desired reactants and products. The methodology used to get thermochemical and kinetic data has been described elsewhere [13]. Thermochemical data for species involved this study have been derived from CBS-QB3 calculations to determine enthalpies of formation, entropies and heat capacities. Explicit treatment of the internals rotors has been performed with the *hinderedRotor* option of Gaussian03 [14]. It must be stressed that in transition states, the constrained torsions of the cyclic structures have been treated as harmonic oscillators and the free alkyl groups as hindered rotations. Enthalpies of formation ($\Delta_f H°$) have been calculated using isodesmic reactions. Rate constants for each elementary reaction were calculated using TST. Tunnelling effect has been taken into account for H transfer processes by using a transmission coefficient as proposed by Wigner [15]. The enthalpies of activation involved in TST theory were calculated by taking into account the enthalpies of reaction calculated with isodesmic reactions in the activation energy. The kinetic data are obtained by fitting the equation of TST at several temperatures between 500 and 2000 K with a modified Arrhenius form:

$$k = A\, T^n \exp(-E/RT) \qquad \textbf{(1)}$$

*ii. Unimolecular initiation of norbornane*

Unlike linear and branched alkanes for which two free radicals are directly obtained, unimolecular initiations of polycyclic alkanes by the breaking of a C-C bond lead to the formation of diradicals (species with two radical centers). The molecule of norbornane (bicyclic alkane) has three different C-C bonds. The unimolecular initiations can lead to the formation of the three diradicals BR1, BR2, and BR3 shown in Figure 3.

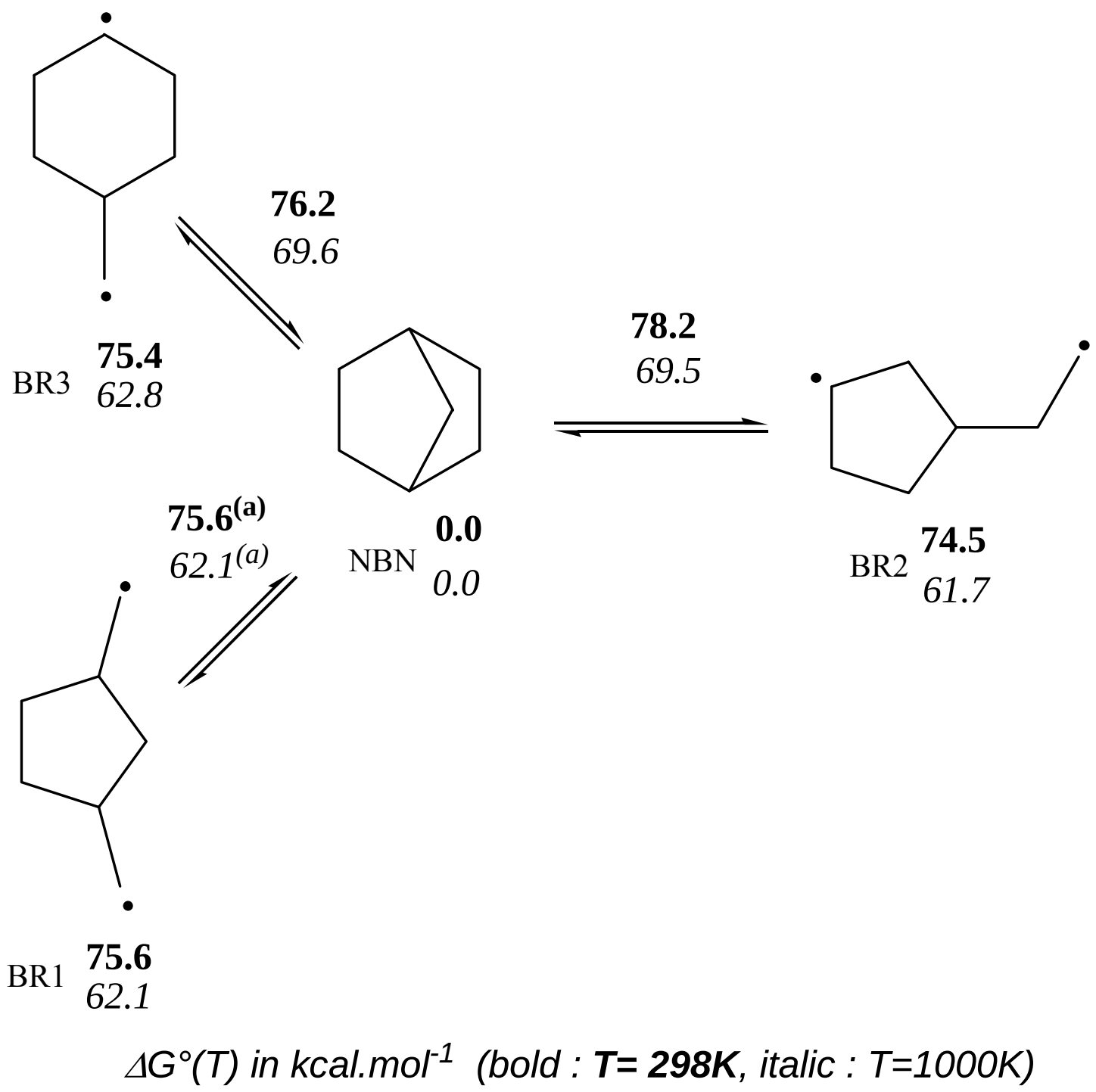

**Figure 3: Unimolecular initiation of norbornane in Gibbs free energies ΔG°(T) at the CBS-QB3 level of calculation. [(a)] Estimated for a barrierless recombination BR1 → NBN (see text).**

Two transition states (TS) have been identified at the B3LYP/cbsb7 level of calculation for the reactions NBN ⇆ BR2 and NBN ⇆ BR3. However, no TS has been charaterized for the reaction NBN ⇆ BR1 on the potential energy surface. The recombination reaction BR1→NBN is a barrierless process, which has been confirmed by a relaxed scan of the potential energy surface.

The rate constants obtained in our calculation are presented in Table 1. Since no TS have been identified for the reaction NBN ⇆ BR1, the rate constant for this process has been estimated using thermokinetics considerations and the relationship proposed by O'Neal [23]:

$$A = e^1 \frac{k_B T}{h} \times rpd \times \exp\left[\frac{3{,}5 \times \Delta n^{\ddagger}}{R}\right] \quad \textbf{(2)}$$

where $k_B$ is the Boltzmann constant, $h$ the Planck constant, $T$ the temperature, $R$ the ideal gas constant, *rpd* the reaction path degeneracy and $\Delta n^{\ddagger}$ the variation of internal rotation between the reactant and the TS.

**Table 1: Kinetic parameters for the unimolecular initiation of NBN. P = 1 atm and 500K < T < 2000K**

| | $k_{NBN\text{-}BR1}$ | $k_{NBN\text{-}BR2}$ | $k_{NBN\text{-}BR3}$ |
|---|---|---|---|
| Log A ($s^{-1}$) | 13,86 | 14,29 | 13,64 |
| n | 0,947 | 0,835 | 0,754 |
| E (kcal/mol) | 81,29 | 81,79 | 78,97 |

In the literature, no rate constant is available for the unimolecular initiation of norbornane. However, it is possible to compare our calculated activation energies ($E_a$) to those obtained using estimations based on a structure–reactivity relationship:

$$E_a = E_{BD} - \Delta E_{RSE} \quad \textbf{(3)}$$

where $E_{BD}$ is the bond dissociation energy of a corresponding linear or branched alkane and $\Delta E_{RSE}$ is the variation of the ring strain energy (RSE) between the reactant (norbornane, RSE = 16.2 kcal/mol according to our calculation) and the transition state. More details on this method can be found in reference [5]. A comparison between the activation energies for the unimolecular initiation of norbornane obtained with equation **(3)** and our calculation is presented in Table 2.

**Table 2: Comparison between the $E_a$ for the unimolecular initiation of norbornane calculated using equation (3) and at the CBS-QB3 level. $E_{BD}$ and RSE have been calculated at the CBS-QB3 level of calculation using isodesmic reactions. Values in kcal/mol, at 298K.**

| Reaction | $E_a$(CBS-QB3) | $\Delta E_{RSE}$ | $E_a$ equation **(3)** | Remaining RSE in the TS |
|---|---|---|---|---|
| NBN ⇆ BR1 | 80.8 | 16.2 – 7.5 = 8.7 | 88.9 – 8.7 = 80.2 | 0.6[(a)] |
| NBN ⇆ BR2 | 81.4 | 16.2 – 7.5 = 8.7 | 86.9 – 8.7 = 78.2 | 3.2 |
| NBN ⇆ BR3 | 78.5 | 16.2 – 1.0 = 15.2 | 86.9 – 15.2 = 71.7 | 6.8 |

[(a)] No TS has been identified for this reaction

From Table 2, it can be seen that a large difference in energy can be observed between $E_a$ (CBS-QB3) and $E_a$ equation **(3)** as large as 6.8 kcal/mol for the reaction NBN $\leftrightarrows$ BR3. This difference can be explained by the fact that equation **(3)** assumes that all the ring strain energy disappears in the transition state as one ring of the bicyclic structure is opened. According to the results presented in Table 2, we can see that this assumption can be erroneous and, consequently, the difference between $E_a$ (CBS-QB3) and $E_a$ equation **(3)** can be defined as the remaining RSE in the transition state. An examination of the geometry of the transition states highlight the reason why the remaining RSE can be so high in the TS. Figure 4 shows the geometry of the TS of the reaction NBN $\leftrightarrows$ BR3 obtained in our calculations.

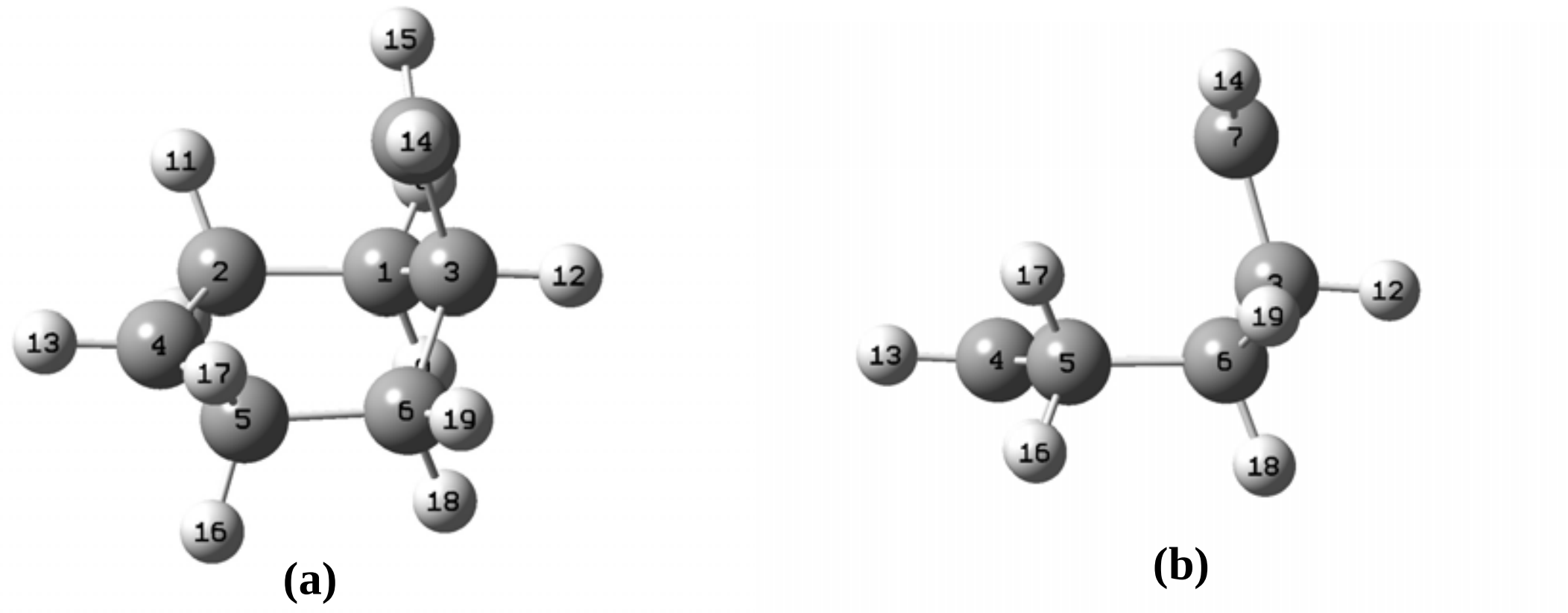

**Figure 4: Geometry of the TS of the reaction NBN $\leftrightarrows$ BR3 obtained at the B3LYP/cbsb7 level of calculation. (a) ¾ top side view (b) side view**

The amount of RSE remaining in the TS can be explained by the fact that a cyclic hydrocarbon structure exists in the TS. In the case of the TS showed in Figure 4 the carbon atoms number 1, 2, 3, 4, 5 and 6 form a cyclohexane-like ring. In this $C_6$ ring, the atoms 1, 2, 4, 5 and 6 belong to the same plan. That conformation of the ring is intermediate between the *chair* (RSE = 1.1 kcal/mol) and *boat* (RSE = 7.5 kcal/mol) conformation of cyclohexane. This conformation is unfavored energetically and the high value of RSE remaining in the TS is principally explained by this steric inhibition.

The same effect is observed in the case of the reaction NBN $\leftrightarrows$ BR2 for which a remaining RSE of 3.2 kcal/mol is estimated in the TS.

The study of the unimolecular initiation of norbornane, which constitute a model polycyclic alkane, underline the considerable contribution of quantum chemistry methods in the study of these hydrocarbons for which the current correlations between structure and reactivity do not apply anymore.

iii. *Fate of diradicals BR1, BR2 and BR3*

The initially formed diradicals can react either by C-C bond breaking by β-scission or by internal disproportionation. In this work we have calculated the activation energies of all the possible pathways of disproportionation of the three diradicals at the CBS-QB3 level. As the activation energy for a C-C bond β-scission (about 28.7 kcal/mol for a linear alkane [16]) is generally higher than that for the internal disproportionation (structure-reactivity correlation leads to $E_a$ around 10 kcal/mol, see Table 3), we only have determined the potential energy surface for the later type of reaction. Figure 5 presents the decomposition scheme of the diradicals formed by the unimolecular initiation of norbornane obtained in our calculations.

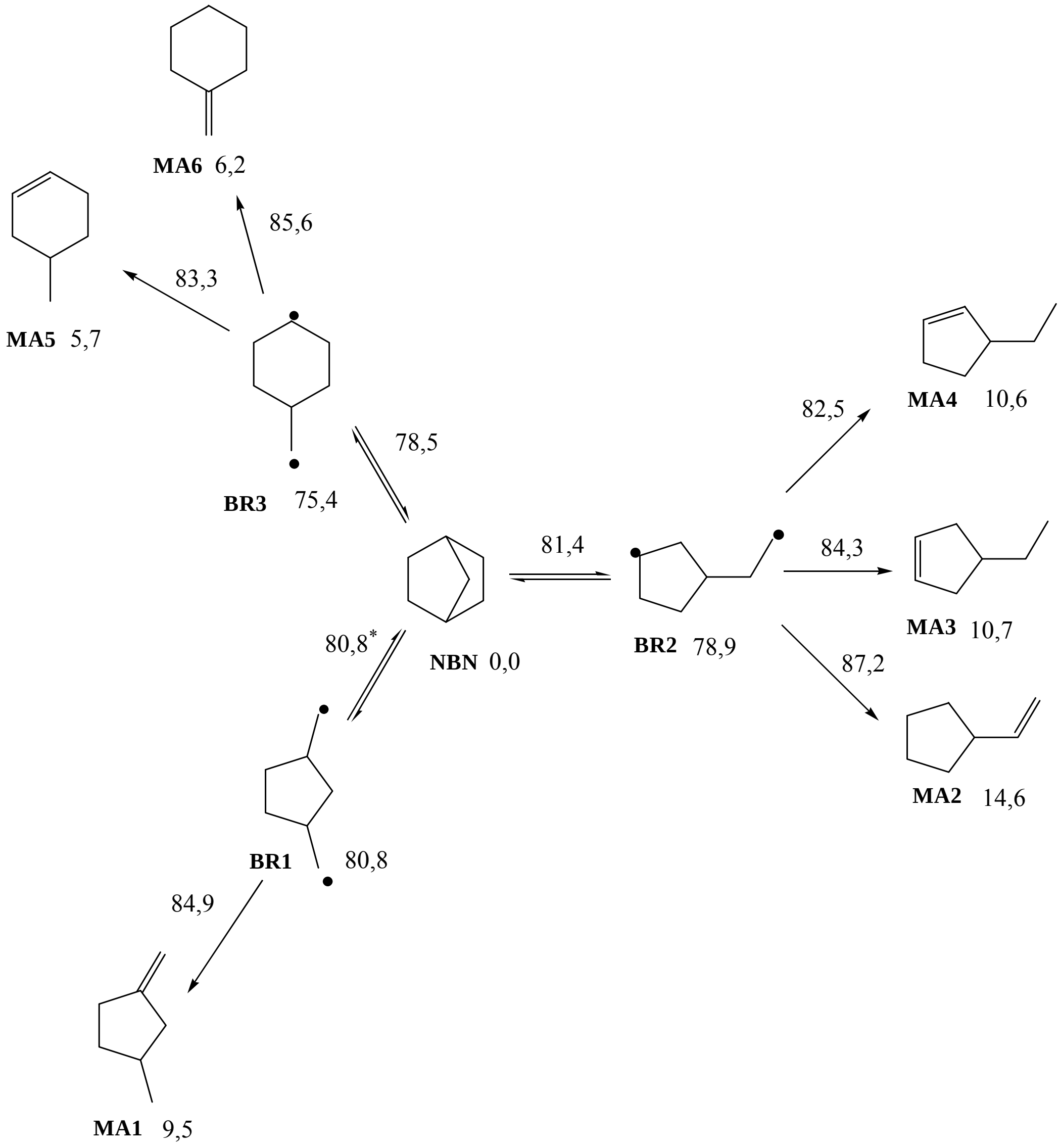

**Figure 5: Fate of diradicals formed by the unimolecular initiation of norbornane. ΔH in kcal/mol at 298 K, referring to norbornane (NBN). * Estimated for a barrierless recombination BR1→NBN.**

From Figure 5, it can be seen that all the $E_a$ for the internal disproportionation are low compared to C-C bond breaking processes, ranging from 3.6 kcal/mol for the reaction BR2 → MA4 to 10.1 kcal/mol for BR3 → MA6. These reactions imply bicyclic transition states for which the associated activation energies cannot be estimate anymore with alkane-based correlations between structure and reactivity. Table 3 shows the kinetic parameters determined for the disproportionation of the diradicals formed during the thermal decomposition of norbornane. The activation energies of the rate constants are those from Figure 5, and the preexponential factors have been estimated using equation **(2)**.

**Table 3: Kinetic parameters for the disproportionation of diradicals BR1, BR2 and BR3. P = 1 atm and 500K < T < 2000K.**

| Réaction | A | n | E |
|---|---|---|---|
| BR1 → MA1 | $3{,}20.10^{9}$ | 1 | 4.1 |
| BR2 → MA2 | $3{,}30.10^{9}$ | 1 | 8,2 |
| BR2 → MA3 | $3{,}30.10^{9}$ | 1 | 5,4 |
| BR2 → MA4 | $3{,}30.10^{9}$ | 1 | 3,6 |
| BR3 → MA5 | $3{,}8.10^{10}$ | 1 | 7,8 |
| BR3 → MA6 | $1{,}9.10^{10}$ | 1 | 10,1 |

iv. *Transfer and propagation reactions of norbornyl radicals*

❖ Decomposition by β-scission of norbornyl radicals

The molecule of norbornane has three different carbon atoms. Reactions of metathesis of hydrogen atoms and radicals with norbornane lead to the formation of three norbornyl radicals. The three norbornyl radicals can react by decompositions by β-scission to yield to the formation of six cyclic radicals (Figure 6). These six new radicals can then react by decompositions by β-scission, by isomerizations, and by metatheses (H abstractions) with molecules. In their work, Herbinet et al. [8] have emphasized the considerable uncertainty related to the activation energy for the ring opening reaction of norbornyl radicals by β-scission. Hence, even for simplest monocyclic alkyl radicals such as cyclopentyl and cyclohexyl it is very difficult to estimate the activation energies. It has been showed that the values used for linear and branched alkyl radicals estimations of activation energies of the reactions of β–scission cannot be used in a systematic way for cycloalkyl radicals [5]. The results of our CBS-QB3 calculations, presented on Figure 6, support this assumption.

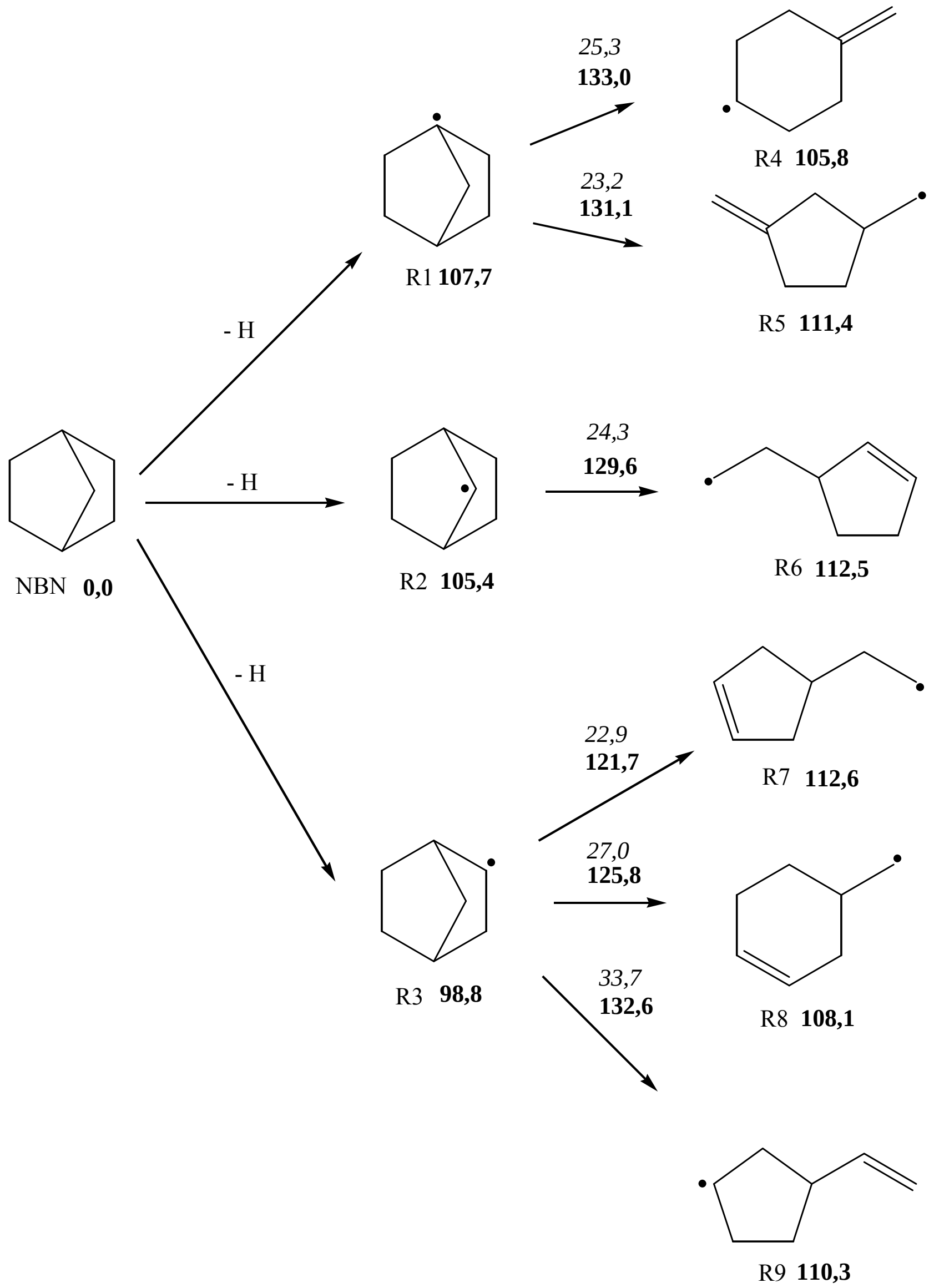

**Figure 6: Decomposition scheme of norbornyl radicals by β-scission of the C-C bonds. All values are enthalpies calculated at the CBS-QB3 level in kcal/mol at 298 K. Values in bold are relative to norbornane, values in italic are relative to the considered norbornyl radical.**

From Figure 6, the bond dissociation energies (BDE) of the three C-H bonds of norbornane can be observed. In the literature, the value of the BDE corresponding to the formation of radical R1 has been proposed by O'Neal et al. [23]. These authors proposed a value of 97.7 kcal/mol for this C-H bond, which is 11 kcal/mol lower than that obtained at our level of calculation. The value of O'Neal et al. is somehow surprising since it is close to the value of 95.7 kcal/mol tabulated for the tertiary H atom of isobutene [24] which is a strain free structure. Figure 7 compares the geometry of norbornane and radical R1 obtained at the B3LYP/cbsb7 level of calculation.

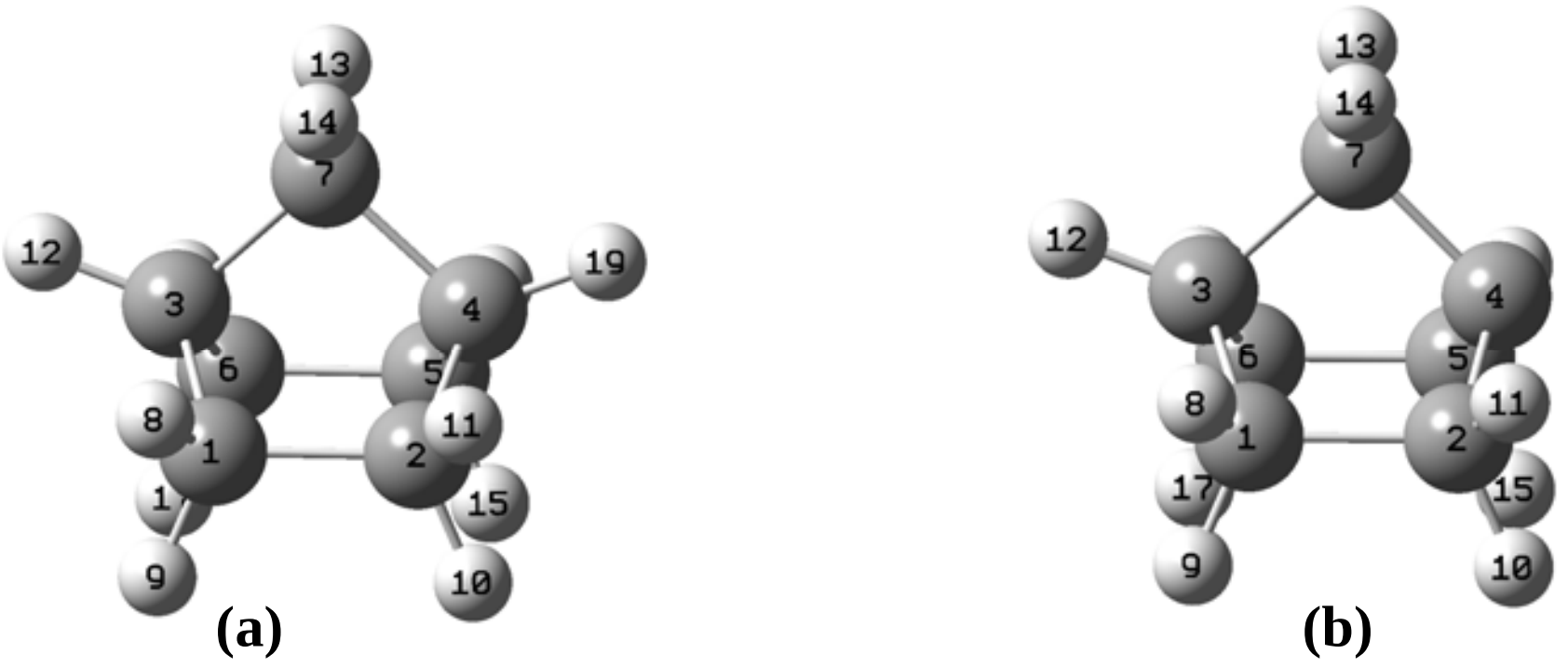

**Figure 7: Geometries of (a) norbornane molecule and (b) norbornyl radical R1 abtained at the B3LYP/cbsb7 level of calculation.**

From Figure 7, it is obvious that the abstraction of a H atom from norbornane leads to a deformation of the bicyclic structure in R1 radical. This deformation is characterized by the fact that the carbon atom number 4 tends to belong to the same plan that the carbon atoms number 2, 5, and 7. Therefore, it is not surprising that the C-H bond dissociation energy is as high as 107.7 kcal/mol since its involved an unfavourable conformation of the bicyclic structure. The same effect is observed for the R2 radical for which one could estimate the C-H BDE to be inferior at 100 kcal/mol (C-H BDE of cyclohexane) and in a less extend for the radical R3. In all these cases the deformation of the polycyclic structure caused by the H abstraction leads to higher C-H BDE. Again, these observations highlight the dramatic lack of data (not only kinetic but also thermodynamic) for the polycyclic hydrocarbons.
The kinetic parameters from our calculations for the unimolecular decomposition pathways of norbornyl radicals are presented in Table 4.

**Table 4: Rate constants for the decomposition of norbornyl radicals.**
**P = 1 atm and 500 K < T < 2000 K.**

| | Log A ($s^{-1}$) | n | E (kcal/mol) |
|---|---|---|---|
| k(**R1** → **R4**) | 12,83 | 0,366 | 26,00 |
| k(**R1** → **R5**) | 12,63 | 0,511 | 23,97 |
| k(**R2** → **R6**) | 12,87 | 0,492 | 24,94 |
| k(**R3** → **R7**) | 12,68 | 0,380 | 23,74 |
| k(**R3** → **R8**) | 12,77 | 0,270 | 27,70 |
| k(**R3** → **R9**) | 12,19 | 0,397 | 34,48 |

In the literature, no rate constant is available for the unimolecular decomposition of norbornyl radicals so that no comparison is possible. The estimation of the activation energies for these processes using structure-reactivity correlations established for branched and linear alkanes highlight the need of new rules, as the alkane-based ones do not apply anymore [5,8].

❖ Unimolecular decomposition of radical R7

The pathway leading to radical R7 is the most favored one in energy according to our calculations (Figure 6). Moreover, sensivity analysis of the kinetic model of the pyrolysis of norbornane have showed that the decomposition pathways of R7 by β-scission play a significant role on the conversion of norbornane as well as on the selectivity of the products. For that reason we have examined the various decomposition pathways of R7 radical at the CBS-QB3 level of calculation. The Figure 8 presents the decomposition scheme obtained for this radical.

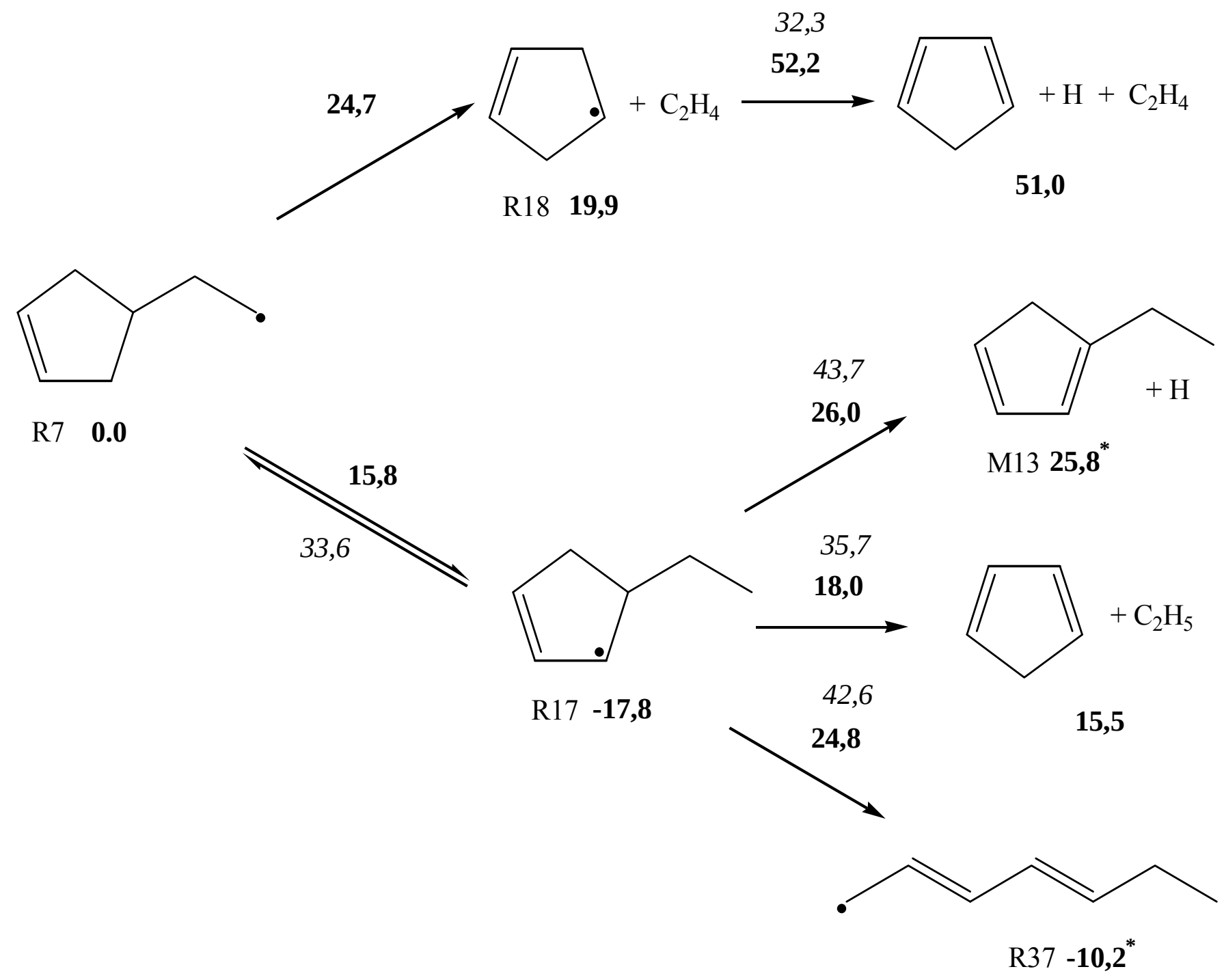

**Figure 8: Unimolecular decomposition of R7 radical. All values are enthalpies calculated at the CBS-QB3 level in kcal/mol at 298 K. Values in bold are relative to R7, values in italic are relative to respectively R17 and R18.**

From Figure 8, it can be seen that the reaction pathways leading to cyclopentadiene are the most favoured energetically via successive β-scissions (R7 → R18 → cyclopentadiene) or via successive isomerization and β-scissions (R7 → R17 → cyclopentadiene). It is worth nothing noting that these most favoured pathways lead to cyclopentadiene and ethylene, two of the three most important pyrolysis products detected in Herbinet et al. experiments [8]. The rate constants for the decomposition of radical R17 have been estimated using equation **(2)** for the preexponential factors A, and the activation energies from our quantum calculations presented on Figure 8. The kinetic parameters from our calculations for the unimolecular decomposition pathways of R7 radicals are presented in Table 5.

**Table 5: Rate constants for the decomposition of R17 radicals. P= 1 atm and 500 K < T < 2000 K.**

| Reaction | $A(s^{-1})$ | n | E(kcal/mol) |
|---|---|---|---|
| R7 → R18 + $C_2H_4$ | $2{,}00.10^{13}$ | 0 | 24,7 |
| R18 → CPD + H | $6{,}40.10^{13}$ | 0 | 32,3 |
| R7 → R17 | $6{,}60.10^{9}$ | 1 | 15,8 |
| R17 → M13 + H | $3{,}20.10^{13}$ | 0 | 43,7 |
| R17 → CPD + $C_2H_5$ | $1{,}30.10^{13}$ | 0 | 35,7 |
| R17 → R37 | $2{,}00.10^{13}$ | 0 | 42,6 |

❖ H-abstraction reactions

Because of the high C-H bond dissociation energies calculated for the norbornane molecule that lead to considerably higher enthalpy of formation of norbornyl radicals compared to the case of unstrained alkanes, we have estimated the rate constants for the H-abstraction reactions by H and $CH_3$ radicals from norbornane using the Evans-Polanyi correlation proposed by Dean and Bozzelli [25]. The rate constants determined are presented in Table 6.

**Table 6: Rate constants for the H-abstraction reactions by H and $CH_3$ radicals from norbornane. Units: $cm^3$, s, kcal, mol.**

| Reaction | A | n | E |
|---|---|---|---|
| **NBN + H → R1 + $H_2$** | $4{,}8.10^{8}$ | 1,5 | 11,7 |
| **NBN + H → R2 + $H_2$** | $4{,}8.10^{8}$ | 1,5 | 7,14 |
| **NBN + H → R3 + $H_2$** | $1{,}92.10^{9}$ | 1,5 | 5,75 |
| **NBN + $CH_3$ → R1 + $CH_4$** | $1{,}62.10^{6}$ | 1,87 | 14,76 |
| **NBN + $CH_3$ → R2 + $CH_4$** | $1{,}62.10^{6}$ | 1,87 | 13,0 |
| **NBN + $CH_3$ → R3 + $CH_4$** | $6{,}48.10^{6}$ | 1,87 | 8,78 |

H-abstraction reactions by cyclopentadienyl radicals from norbornane have also been added to the mechanism using the kinetic parameters proposed by Buda et al. [22].

### *c. Cross-Coupling Reactions between benzene and norbornane mechanisms*

As norbornane is solid at room temperature, it has been dissolved in benzene in the study of of the thermal decomposition of norbornane of Herbinet et al. [5]. These authors have showed experimentally that there are low interactions between the two hydrocarbons. In our kinetic model, we have taken into account the H-abstraction reactions by H, $CH_3$, $C_2H_5$, and $aC_3H_5$ (allyl radical) on benzene leading to the formation of phenyl radical as well as the termination reaction of these small radicals with phenyl radicals. We have also added the H-abstraction reactions of phenyl radical on the norbornane molecule. The kinetic parameters for these processes have been taken from the work of Ziegler et al. [26].

### *d. Consumption of molecular products*

The decomposition reactions of some important molecular products have been added to the detailed mechanism. The decomposition mechanisms of cyclopentadiene, cyclohexadiene, toluene, styrene and biphenyl have been taken from the oxidation kinetic model of toluene proposed by Bounaceur et al. [27].

## 2. Simulation results and discussion

Simulations were performed using PSR of CHEMKIN II software [28] and the mechanism described above which involves 209 species and includes 497 reactions. The agreement between computed and experimental [5] mole fractions vs residence time is globally satisfactory for conversion of norbornane (Figure 9) at 933 K, 953 K and 973 K, and for products such as hydrogen, ethylene, 1,3-cyclopentadiene, methane, propene, ethane, (Figure 10) and larger species such as 1,3-butadiene, methylcyclopentadiene, biphenyl and toluene (Figure 11).

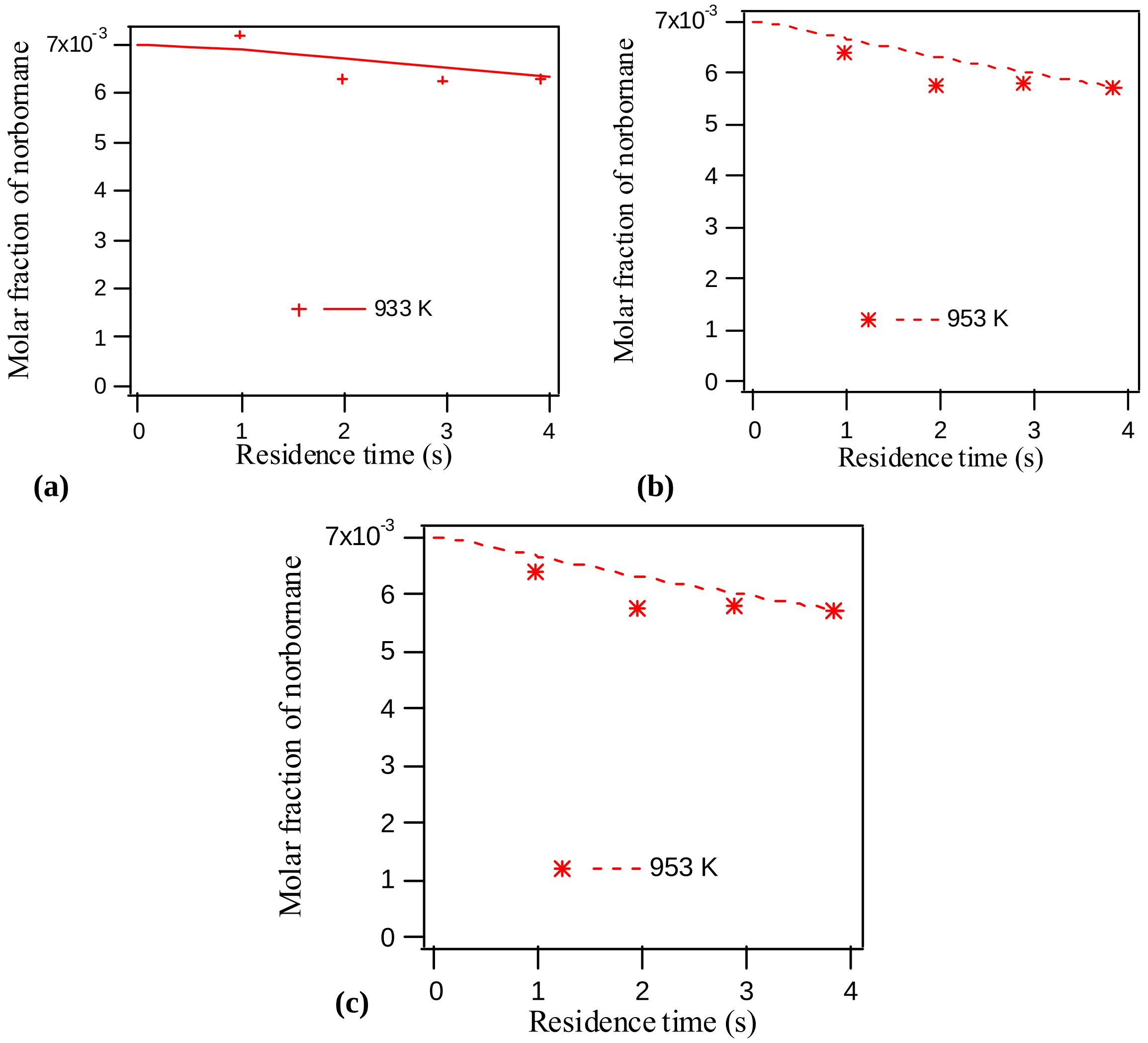

**Figure 9: Mole fractions of norbornane vs residence time at several temperatures for an initial norbornane mole fraction of 0.7%.**

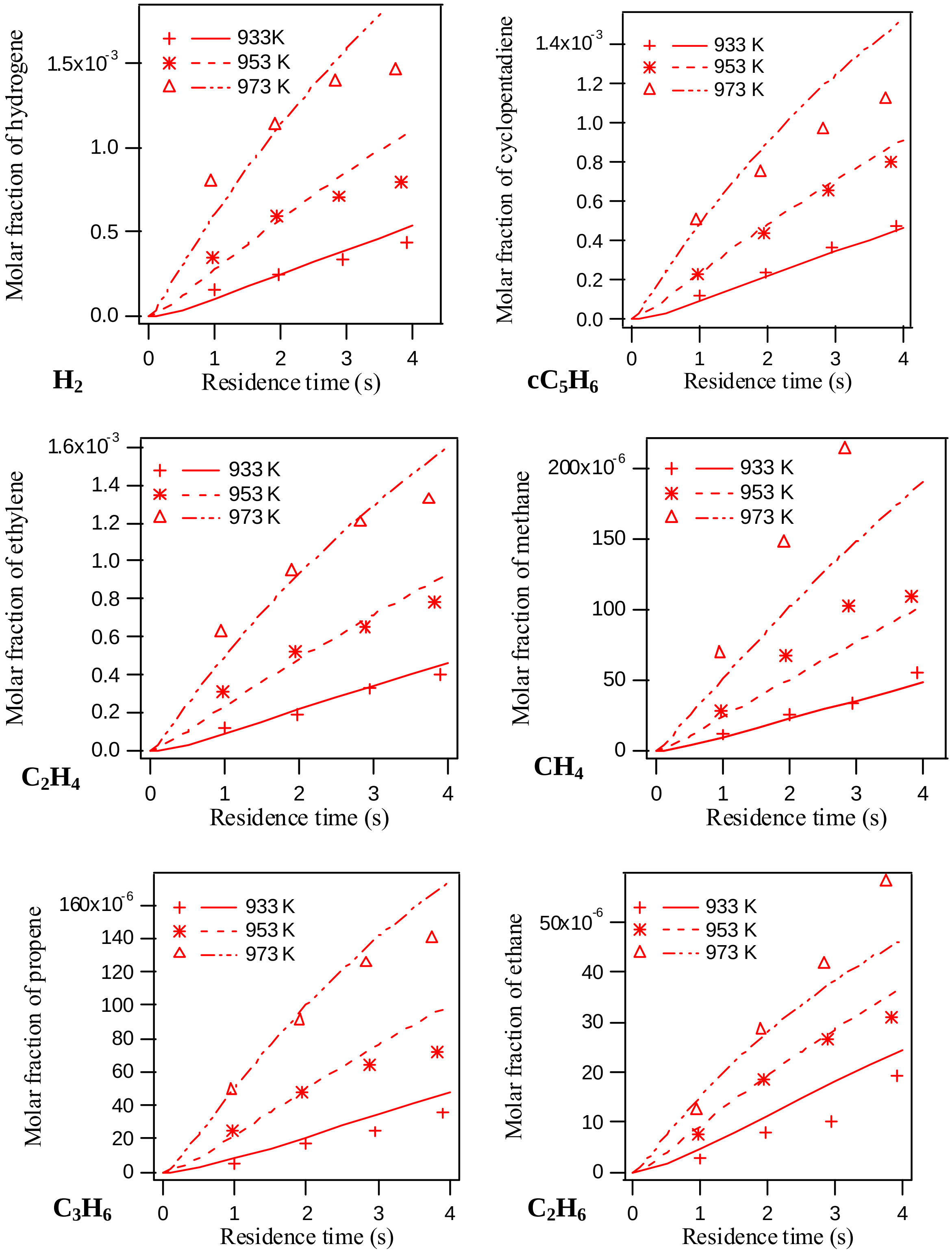

**Figure 10: Mole fractions of light products vs residence time at several temperatures for an initial norbornane mole fraction of 0.7%.**

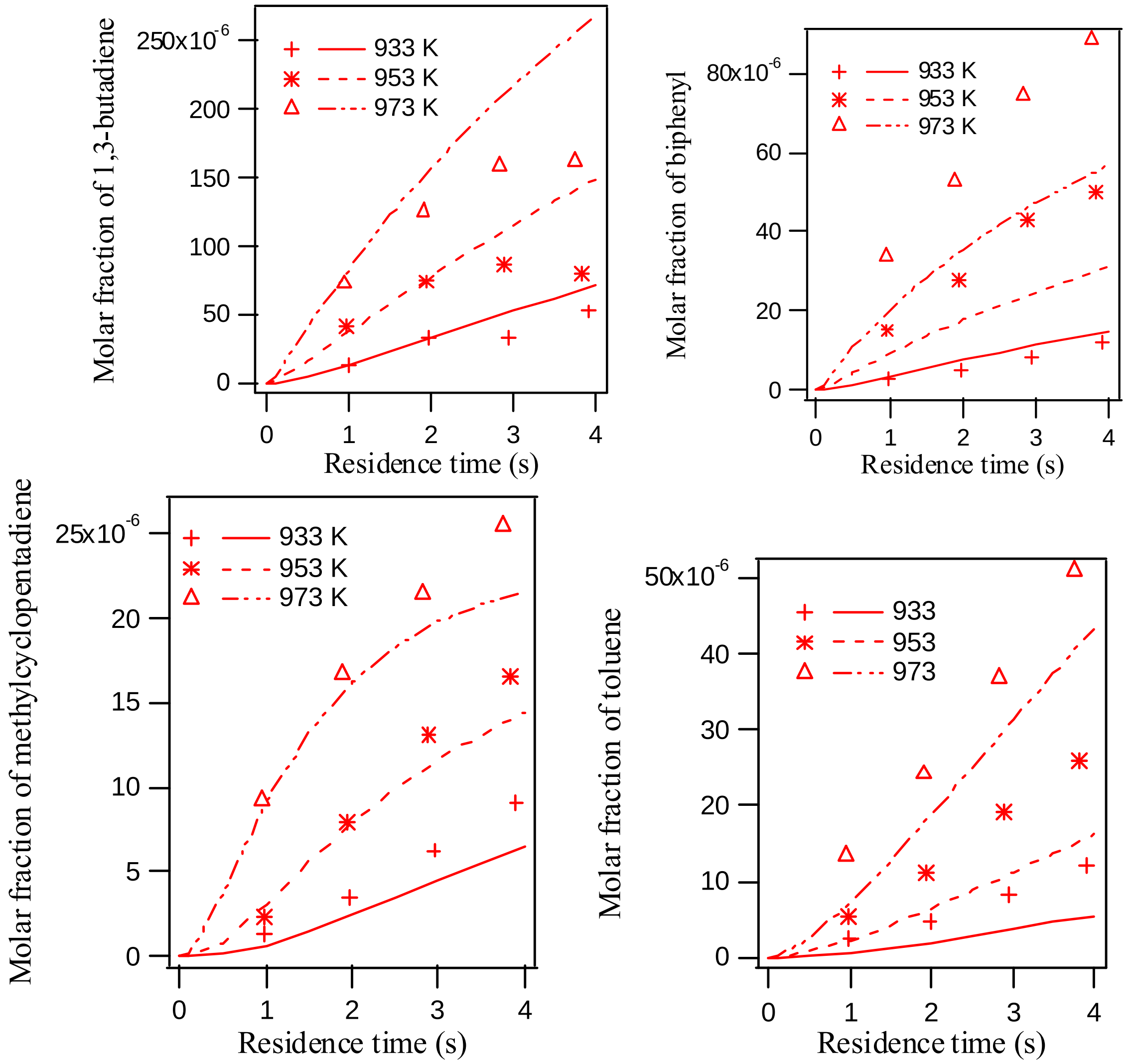

**Figure 11: Mole fractions of heavy products vs residence time at several temperatures for an initial norbornane mole fraction of 0.7%.**

Figure 12 presents a flow rate analysis performed at 933 K and at a residence time of 1 s corresponding to a conversion of 4.4% of the norbornane. The arrows and the percentages represent the flow of consumption of a species through a given channel. Flow rate analysis shows that 14.1% of the reactant is consumed through unimolecular initations which underline the importance of unimolecular initiations. 8.2% of the initiation flow lead to BR2 diradical and 5.9% lead to BR3 diradical. These two diradicals react by internal disproportionation leading to primary molecules that in turn decompose by unimolecular initiations leading to species that will be either radicals or molecules, leading to different consequences on the reactivity of the system.

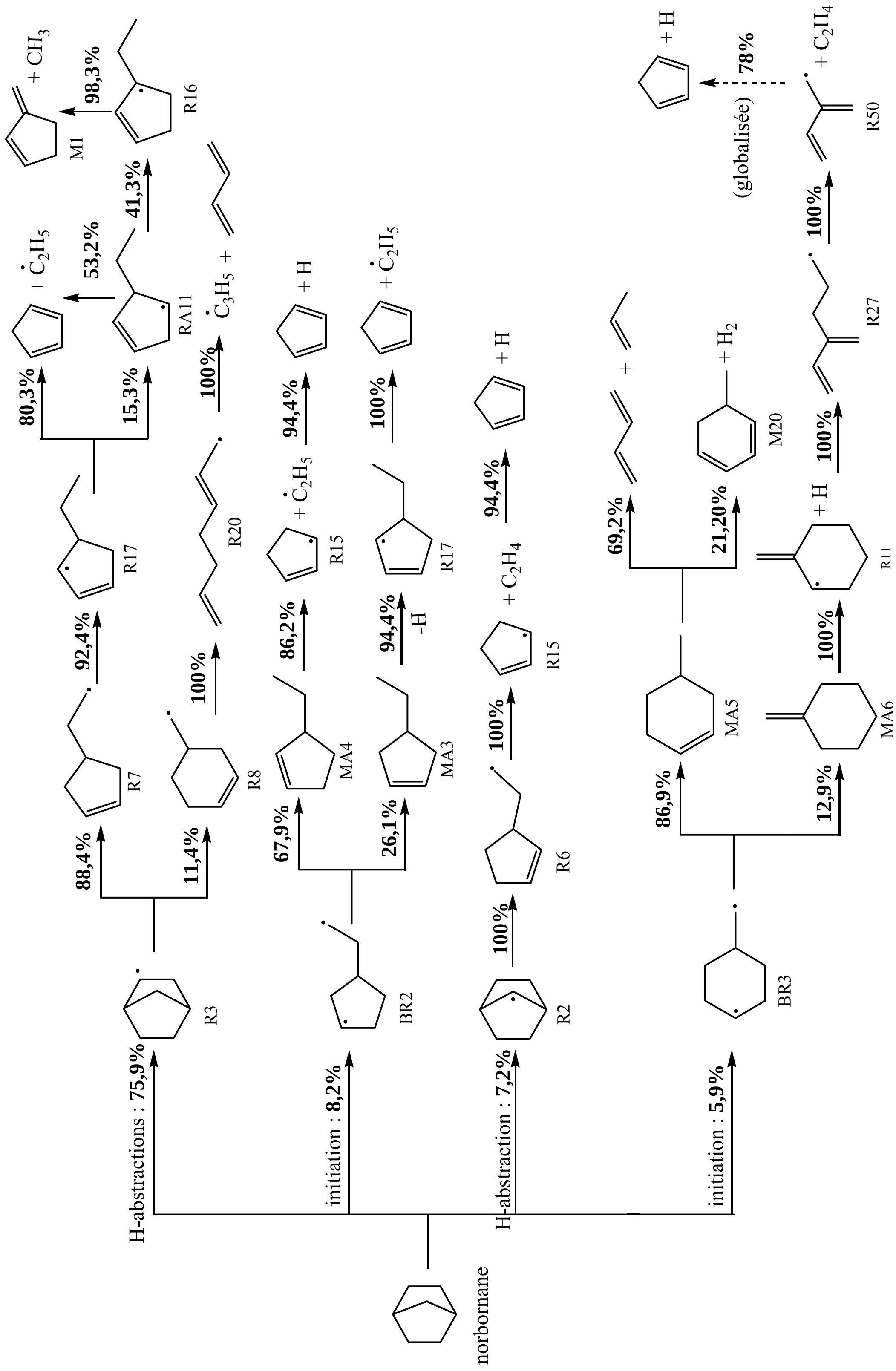

**Figure 12: Major flows of consumption of tricyclodecane at a temperature of 953 K, a residence time of 1 s, and an initial hydrocarbon mole fraction of 0.7% (conversion of 4.4%).**

The major channel of decomposition of BR3 diradical leads to MA5 molecule (4-methylcyclohexane) which decomposes by molecular reaction leading to low reactive species such as 1,3-butadiene and propene. This pathway is very important for the product selectivity as it contributes to 33.5% of the total product flow of propene and 20% of the total product flow of 1,3-butadiene.

The major channel of decomposition of BR2 diradical leads to ethylcyclopentene molecules (MA3 and MA4) by internal disproportionation. The major decomposition channel leads to cyclopentenyl (R15) and ethyl radicals via MA4. These two radicals can in turn lead to cyclopentadiene and ethylene by H-atom release. The H atom production is favored by this pathway which have a strong promoting effect on the reactivity.

Norbornane is mainly consumed by H-abstraction reactions with small radicals such as H (67.6% of the total H-abstraction flow), $CH_3$ (4.9%) and allyl (3.4%) leading to the formation of only two norbornyl radicals (R3 and R2). A very large part of the H-abstractions lead to R3 radicals (75.9%). This can be explained by the highest stability of this radical (see Figure 6). This fact highlights the critical need of accurate thermodynamic data for the polycyclic alkanes. R3 radicals yield mainly to cyclopentadiene and ethyl radicals via R7 and R17 radicals. This channel is very important as it constitutes 63,7% of the total production flow of cyclopentadiene and 85.5% of the total production flow of ethyl radicals leading to ethylene. Another important path, regarding the selectivity of the products, involves the formation of RA11 from R7 and R17 via internal isomerization leading to the production of methyl radical (63% of the total production flow of methyl) which in turn yield to methane, one of the eight principal products of pyrolysis. The other pathway of decomposition of R3 radical finally leads to 1,3-butadiene (79.9% of the total product flow of this molecule) and allyl radical (99.8% of the total production flow of this radical). The flow rate analysis underline the considerable contribution of quantum chemistry methods to describe key decomposition channels as a few pathways are involved to reproduce the experimental selectivity of the products with simulations. This remark is also supported by the sensitivity analysis on the consumption of norbornane displayed in Figure 13.

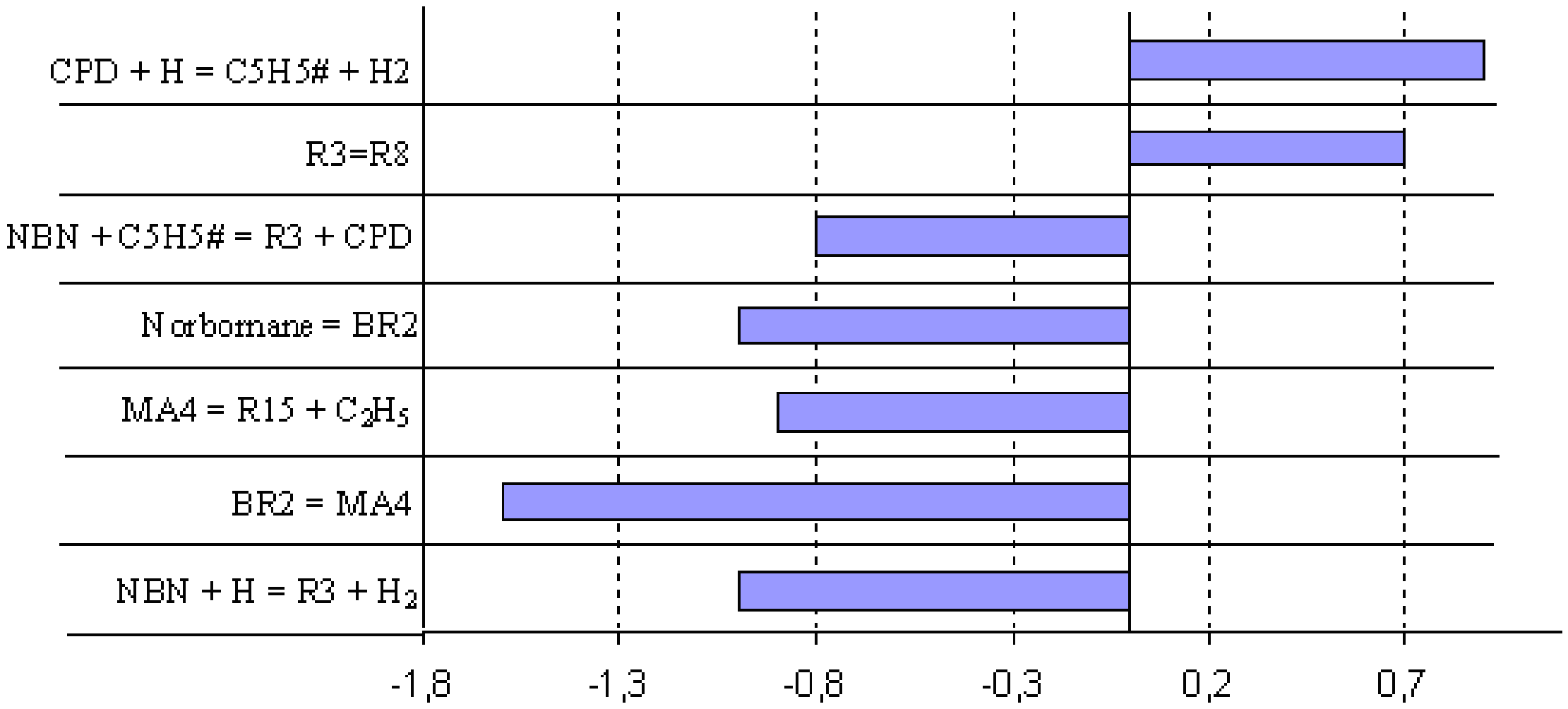

**Figure 13: Sensitivity analysis for norbornane mole fraction (T = 953 K and residence time of 1s).**

Sensitivity analysis on norbornane shows that the conversion of the reactant is strongly controlled by the unimolecular initiation leading to the formation of diradical BR2 (NBN=BR2) and by the internal disproportionation of this diradical yielding to MA4 (BR2=MA4). The unimolecular initiation of MA4 also plays an important role on the consumption of norbornane. The sensitivity analysis shows that these unimolecular initiations have a promoting effect on the conversion of the reactant as they lead to the formation of very reactive H radicals. It is also the case of the H-abstraction reactions by H and cyclopentadienyl radicals from norbornane which lead to R3 radicals whose decomposition mainly yields ethyl radicals.
Two reactions have an important inhibiting effect on the consumption of norbornane: the H-abstraction by H atom from cyclopentadiene leading to cyclopentadienyl and $H_2$ which consumes very reactive H atoms to produce low reactive cyclopentadienyl radicals; and the ring opening reaction of R3 yielding to R8 (that compete with the promoting channel R3 → R7) which decomposition leads to low reactive species such as 1,3-butadiene and resonantly stabilized allyl radicals.

## Conclusion

In this work, a detailed chemical kinetic model for the pyrolysis of norbornane, a model molecule for jet fuel surrogate, has been developed following EXGAS software methodology and validated very satisfactorily against experimental data [5] obtained in a jet-stirred reactor. The sensitive reaction channels, e.g. unimolecular initiation of norbornane, the fate of the formed diradicals, the decomposition of the norbornyl radicals, have been described by quantum chemistry calculation. The kinetic analyses of the mechanism have shown that the unimolecular initiations and the decomposition of the formed diradicals played an important role on the consumption of norbornane. The main flow of consumption of norbornane occurs through H-abstraction reactions leading mainly to the formation of R3 radical. The decomposition of this radical by β-scission and the subsequent unimolecular reactions of the formed radicals (R7 and R8) constitute the major formation pathway of all the species detected experimentally in a large amount: cyclopentadiene, ethylene (from ethyl radical), propene (from allyl radical), methane (from methyl radical) and 1,3-butadiene. The kinetic analysis of the model for the pyrolysis of norbornane highlights the crucial needs of accurate thermodynamic and kinetic parameters for the polycyclic alkanes decomposition reactions. The ring strain energies of (mono or poly) cyclic alkanes do not allow anymore the use of alkane-based correlations between structure and reactivity to estimate the thermokinetic parameters of the reactions of these hydrocarbons. In absence of more experimental data, quantum chemistry is therefore very valuable, not only to calculate accurately these parameters but also to explore all the possible pathways of these very complex structures. A considerable work remains to be done on polycyclic alkanes reaction under combustion condition and quantum chemistry methods will play a central role in the determination of new correlations between structure and reactivity for this important class of hydrocarbon.

## Acknowledgments

The Centre Informatique National de l'Enseignement Supérieur (CINES) is gratefully acknowledged for allocation of computational resources.